\documentstyle[prl,aps]{revtex}

\begin{document}

\topmargin 0pt
\oddsidemargin 5mm

\setcounter{page}{0}
\begin{titlepage}
\vspace{2cm}
\begin{center}

{\Large Interference phenomena in radiation of a charged particle\\ 
moving in a system with one-dimensional randomness}\\
\vspace{1cm}
{\large Zh.S. Gevorkian$^{1}$ and Th.M. Nieuwenhuizen$^{2}$}\\
\vspace{3mm}
{\em $^{1}$Institute of Radiophysics and Electronics}\\
{Ashtarak-2, 378410, Armenia}\\
{\em $^{2}$Van der Waals-Zeeman Instituut, Valckenierstraat 65,\\
1018 XE Amsterdam, The Netherlands\\e-mail: nieuwenh@phys.uva.nl}\\
\vspace{5mm}
{\bf{Abstract}}
\end{center}

The contribution  of interference effects to the radiation of a charged 
particle moving in a medium of randomly spaced plates is considered. 
In the angular dependent radiation intensity a peak appears at
angles $\theta\sim\pi-\gamma^{-1}$, where $\gamma$ is the Lorentz factor
of the charged particle.
\vfill
\end{titlepage}
\section{Introduction}
\indent
It is well known that a
charged particle passing through a stack of randomly spaced 
plates radiates electromagnetic waves (see, for example, [1]). The radiation 
is caused by the scattering of the electromagnetic field
(pseudo-photon) of the charged particle from the 
inhomogeneities in the dielectric constant. In an  earlier study
one of us has shown [2] that, in analogy with three-dimensional 
random media [3], the multiple scattering of the electromagnetic 
field plays an important role. 
However, in the multiple scattering approach only 
the diffusion contribution was taken into account. 
At this level the approach is equivalent to the radiative transfer 
theory for light transport in e.g. slab
geometries, see [4] for a recent review.

On the other hand, interference effects 
are important when waves propagate in random inhomogeneous media. 
Anderson localization [5] and the enhanced backscattering peak [6] are 
manifestations of these effects. Other interference effects show up
in correlations and higher moments of the transmitted intensity.
They were also reviewed in [4].

In the present paper we want to investigate  
interference effects for radiation of a charged particle moving in a 
system with one-dimensional randomness.

\section{Formulation of the Problem}
\indent

The system which we want to study consists of a stack of plates
randomly spaced in a homogeneous medium.  The dielectric constant of the 
system can be represented in the form

\begin{equation}
\label{1}%1
\varepsilon(z,\omega )=\varepsilon  _0(\omega)
+\sum_{i}\left[b(\omega )-\varepsilon
_0(\omega)\right]\left[|\Theta(z-z_i-a/2)-\Theta(z-z_i+a/2)|\right],
\end{equation}
where $z_{i}$  are the random coordinates of the plates, $a$ is their 
thickness, and $\varepsilon_0(\omega)$ and $b(\omega)$ are
die\-lec\-tric permeabilities of the homogeneous me\-di\-um and
the plate, respectively.
It is convenient to represent the dielectric
permeability as the sum of an average and a fluctuating part

\begin{equation}
\label{2}%2
\varepsilon (z,\omega)=\varepsilon+\varepsilon_r(z,\omega),\quad
\quad < \varepsilon _r(z,\omega)>=0,
\end{equation}
where $\varepsilon =<\varepsilon(z,\omega)>$ and
averaging over the random coordinates of  plates is determined as
follows
\begin{equation}
\label{3}%3
<f(z,\omega)>=\int \prod_{i}\frac{dz_i}{L_z}f(z,{z_i},\omega ),
\end{equation}
where $L_z$ is the system size in the $z$-direction.
The vector potential of the electromagnetic field created by a moving charged particle satisfies the equations.

\begin{eqnarray}
\label{4}%4
div\vec{A}-\frac{i\omega}{c}\varepsilon(\vec{r},\omega)
\varphi (\vec{r},\omega)=0 ,\nonumber \\
\nabla^2\vec{A}+\frac{\omega^2}{c^2}
\varepsilon (\vec{r},\omega)\vec{A}(\vec{r},\omega)=\vec{j}
(\vec{r},\omega)
\end{eqnarray}
where $\vec{j}(\vec{r},\omega)$ is the current associated with the moving 
charged particle

\begin{equation}
\label{5}%5
\vec{j}(\vec{r},\omega)=-\frac{4\pi e}{c}\,\frac{\vec{v}}{v}
\delta(x)\delta(y) e^{i\omega z/v}, \vec{v}\|z
\end{equation}
The symmetry of the problem allows us to choose the vector potential along 
the $z$-axis, $A_{i}=\delta_{\hat{z}i}A$. 
The electric field is expressed through 
the potentials in the following way

\begin{equation}
\label{6}%6
\vec{E}(\vec{r},\omega)=\frac{i\omega}{c}\vec{A}(\vec{r},\omega)
-{\rm grad} \varphi (\vec{r},\omega)
\end{equation}
As usually, we decompose electric field into two parts: $\vec{E}=\vec{E_{0}}
+ \vec{E_{r}}$. Here $\vec{E_{0}}$ is the field originated by the charged 
particle moving in the homogeneous medium with dielectric constant 
$\varepsilon$ and  $\vec{E_{r}}$ is the radiation field associated with 
the fluctuations of dielectric constant. The radiation tensor is determined 
as follows

\begin{equation}
\label{7}%7
I_{ij}(\vec{R})=E_{ri}(\vec{R})E_{rj}^*(\vec{R})
\end{equation}
where $\vec{R}$ is the radius-vector of the observation point which is far 
away from the system,  $R\gg L$. For expressing the radiation intensity 
through the radiation potential $\vec{A_{r}}$, we decompose the vector 
potential analogous to the decomposition of electric field $\vec{A}=
\vec{A_{0}}+\vec{A_{r}}$. The fields $\vec{A_{0}}$ and $\vec{A_{r}}$ satisfy 
the equations

\begin{eqnarray}
\label{8}%8
&&\nabla^2\vec{A}_0+\frac{\omega^2}{c^2}\varepsilon\vec{A}_0=
\vec{j}(\vec{r},\omega)\nonumber \\
&&\nabla^2\vec{A}_r+\frac{\omega^2}{c^2}\varepsilon\vec{A}_r+
\frac{\omega^2}{c^2}\varepsilon_r\vec{A}_r=-\frac{\omega^2}{c^2}
\varepsilon_r\vec{A}_0
\end{eqnarray}
Using (\ref{4}) and (\ref{6}) one can express the radiation tensor in terms 
of radiation potential          

\begin{eqnarray}
\label{9}%9
&&<I_{ij}(\vec{R})>=\frac{\omega^2}{c^2}\delta_{\hat{z}i}\delta
_{\hat{z}j}<A_r(\vec{R},\omega)A_r^*(\vec{R},\omega)>
+\frac{\delta_{\hat{z}i}}{\varepsilon}<A_r(\vec{R},\omega)
\frac{\partial^2}{\partial R_j\partial z}
A_r^*(\vec{R},\omega)>\nonumber \\
&&+\frac{\delta_{\hat{z}j}}{\varepsilon}<A_r^*(\vec{R},\omega)
\frac{\partial^2}{\partial R_i\partial z}
A_r(\vec{R},\omega)>+\frac{c^2}{\omega ^2 \varepsilon^2}<
\frac{\partial^2}{\partial R_i\partial z}
A_r(\vec{R},\omega)\frac{\partial^2}{\partial R_j\partial z}
A_r^*(\vec{R},\omega)>
\end{eqnarray}
For obtaining (\ref{9}) we supposed that $\varepsilon_{r}\ll\varepsilon$.
 In order to carry out the averaging over the random coordinates of the 
plates it is convenient to express the radiation potential in terms of the 
Green's function of equation (\ref{8})

\begin{eqnarray}
\label{10}%10
&& A_r(\vec{R})=-\frac{\omega ^2}{c^2} \int \varepsilon_r(\vec{r})
A_0(\vec{r})G(\vec{R},\vec{r})d\vec{r}\nonumber \\
&&\left[ \nabla^2+k^2+\frac{\omega ^2}{c^2} \varepsilon_r(z)\right]
G(\vec{r},\vec{r}{\,}^{\prime})= \delta(\vec{r}-\vec{r}^\prime)
\end{eqnarray}
where $k=\omega\sqrt{\varepsilon}/c$.

\section{Green's  Function}
\indent

In this chapter we follow the approach of our previous paper [2]. For the 
bare Green's function in momentum representation one has from (\ref{10})

\begin{equation}
\label{11}%11
G_0(\vec{q})=\frac{1}{k^2-q^2+i\delta}
\end{equation}
In the coordinate representation one has 

\begin{equation}
\label{12}%12
G_0(r)=-\frac{1}{4\pi  r}e^{ikr}
\end{equation}
The average Green's function in the independent scatterer approximation has 
the form

\begin{equation}
\label{13}%13
G(\vec{q})=\frac{1}{k^2-q^2+i{\rm Im}\Sigma(\vec{q})},
\end{equation}
where the imaginary part of the self-energy ${\rm Im}\Sigma(\vec{q})$ is 
determined self-consistently by the Ward identity

\begin{eqnarray}
\label{14}%14
&&{\rm Im}\Sigma(\vec{q})=\int \frac {d\vec{p}}{(2\pi
)^3}B(\vec{p}){\rm Im}
G_0(\vec{q}-\vec{p})=\frac{1}{4\sqrt{k^2-q^2_\rho }}\nonumber \\
&&\left[ B(|q_z-\sqrt{k^2-q^2_\rho }|)+B (|q_z+\sqrt
{k^2-q^2_\rho }|)\right],\quad |\vec{q}_\rho |<k
\end{eqnarray}
Here $B(\vec{p})=(2\pi)^2\delta(\vec{p}_{\rho})B(|p_{z}|)$, where 
$\vec{p_{\rho}}$ is the transverse component of the $\vec{p}$ and $B(|p_{z}|)$ 
is the correlation function of the one-dimensional random field

\begin{equation}
\label{15}
B(|z-z^\prime|)=\frac{\omega ^4}{c^4}<\varepsilon
_r(z)\varepsilon_r(z^\prime)>
\end{equation}
Using explicit form (\ref{1}) of the random field and carrying out the 
averaging with the help of (\ref{3}), we obtain

\begin{equation}
\label{16}
B(q_z)=\frac{4(b-\varepsilon)^2 n \sin^2q_z
a/2}{q_z^2}\frac{\omega ^4}{c^4}
\end{equation}
where $n$ is the concentration of plates in the system.
  The photon mean free path in the $z$ direction is determined by

\begin{equation}
\label{17}
l(\vec{q})=\frac{\sqrt{k^2-q^2_\rho }}{{\rm Im}\Sigma(\vec{q})}
\end{equation}
As one could expect, the photon mean free path depends on the momentum 
direction. This is principally different from the isotropic case. 
In the case when the momentum is directed along $z$, using (\ref{14}) and 
(\ref{17}), one obtains

\begin{equation}
\label{18}
l(\theta=0 )=\frac{4k^2}{B(0)+B(2k)}
\end{equation}

From this point we shall call this quantity the pseudo-photon mean free path.
It follows from (\ref{16}) that $B(0)=(b-\varepsilon)^{2}\omega^{4}na^{2}/
c^{4}$ and $B(2k)/B(0)\sim\sin^{2}{ak}/a^{2}k^{2}$. Therefore 
we have for the mean free path 

\begin{equation}
\label{19}
l\equiv l(\theta =0)\approx \left\{ \begin {array}{ll}
4k^2/B(0),&\,ka\gg 1\\
2k^2/B(0), &\, ka\ll 1 \end{array} \right.
\end{equation}
Note that the expressions (\ref{19}) are obtained in the Born approximation 
which holds when $|\sqrt{b/\varepsilon}-1|ka\ll1$. Notice also the 
restriction imposed on the angles. As was mentioned in [2] our consideration 
is correct up to angles $\theta\approx\pi/2-\delta$ (where $|\delta|\gg(1/kl)^
{1/3}$). 

\section{The Radiation Intensity}
\indent
Substituting (\ref{10}) into (\ref{9}) we obtain following expression for 
the averaged radiation tensor

\begin{eqnarray}
\label{20}
<I_{ij}(\vec{R})>=\int d\vec{r}d\vec{r{\,}^{\prime}}A_{0}(\vec{r})
A^{*}_{0}(\vec{r{\,}^{\prime}})<\varepsilon_{r}(\vec{r})
\varepsilon_{r}(\vec{r{\,}^{\prime}})\nonumber \\
\left[\delta_{\hat{z}i}\delta_{\hat{z}j}
\frac{\omega^{6}}{c^{6}}G(\vec{R,r})G^{*}(\vec{r{\,}^{\prime},R})+
\frac{\omega^{2}}
{c^{2}\varepsilon^{2}}\frac{\partial^2G}{\partial R_i\partial z}
\frac{\partial^2G^*}{\partial R_j\partial z}
+\frac{\delta_{\hat{z}j}}{\varepsilon}\frac{\omega^{4}}{c^{4}}G^*
\frac{\partial^2G}{\partial R_i\partial z}+
\frac{\delta_{\hat{z}i}}{\varepsilon}\frac{\omega^{4}}{c^{4}}G
\frac{\partial^2G^*}{\partial R_j\partial z}\right]>
\end{eqnarray}
As mentioned above, the observation point is far away from the 
radiating system. For this reason, using (\ref{12}), one can obtain following 
useful relations

\begin{equation}
\label{21}
G_0(\vec{R},\vec{r})\approx-\frac{1}{4\pi
R}e^{ik(R-\vec{n}\vec{r})},\,\frac{\partial^2 G_0(\vec{R},\vec{r})}
{\partial R_i\partial z}\approx \frac{k^2n_in_z}{4\pi R}
e^{ik(R-\vec{n}\vec{r})},\,R\gg r
\end{equation}
Here $\vec{n}$ is the unit vector in the direction of the observation point 
$\vec{R}$. The radiation tensor (\ref{20}) consists of three contributions. 
Single scattering and diffusion contributions have been studied in a previous 
paper [2]. Therefore in the present paper we shall give our main attention to 
the interference contribution.
 Substituting (\ref{21}) into (\ref{20}) and using (\ref{15}) we obtain for 
the single scattering contribution

\begin{eqnarray}
\label{22}
&& I^0_{ij}(\vec{R})=\frac{k ^2}{16\pi^2 R^2\varepsilon}\,
\int d\vec{r}d\vec{r}{\,}^\prime
B(r-r^\prime) A_0(\vec{r})A_0^*(\vec{r}{\,}^\prime)\,e^{-ik\vec{n}(\vec{r}-
\vec{r}{\,}^{\prime})}\nonumber\\
&&\left[ \delta  _{\hat{z}i}\delta  _{\hat{z}j}+
n_i n_j n_z^2-\delta _{\hat{z}i}n_jn_z-
\delta_{\hat{z}j}n_i n_z\right]
\end{eqnarray}
Solving (\ref{8}), one obtains for the background potential

\begin{equation}
\label{23}
A_0(\vec{q})=-\frac{8\pi^2 e}{c}\,\frac{\delta  (q_z-\omega
/v)}{k^2-q^2}
\end{equation}
Substituting (\ref{23}) into (\ref{22}) after some straightforward 
calculations we find for the single scattering contribution to the 
radiation intensity $I=\frac{cR^2}{2}I_{ii}(\vec{R})$ [2],

\begin{equation}
\label{24}
I^0(\theta )=\frac{e^2}{2c}\frac{L_z B(|k_0-k\cos\theta
|)\sin^2\theta }{{\left[\gamma ^{-2}+\sin^2\theta\,k^2/k^2_0
\right]}^2}\,\frac{\omega ^2}{k_0^4 c^2}
\end{equation}
where $\gamma ={(1-\varepsilon v^2/c^2)}^{-1/2}$ is the Lorentz factor of 
the relativistic particle in the medium, $k_{0}=\frac{\omega}{v}$,
$n_{z}=\cos{\theta}$ and $L_{z}$ is the system size in the $z$ direction. 
As seen from (\ref{24}) at $ak\ll1$ ($B\sim const$), the forward and 
backward intensities are equal. When $ak\gg1$, for relativistic particles
$k_{0}\rightarrow k$, $\gamma\gg 1$ the forward intensity ($\theta\approx 0$) 
is significantly larger than the backward intensity ($\theta\approx\pi$) 
because of the factor $B$.
  The diffusion contribution can be represented in the form [2]

\begin{eqnarray}
\label{25}
&& I^D_{ij}(\vec{R})=\frac{k ^2}{16\pi^2 R^2\varepsilon}\,
\int d\vec{r}d\vec{r}{\,}^\prime
B(r-r^\prime) A_0(\vec{r})A_0^*(\vec{r}{\,}^\prime)\,\int
d\vec{r}_1d\vec{r}_2 d\vec{r}_3 d\vec{r}_4\nonumber \\
&&e^{-ik\vec{n}(\vec{r}_1-\vec{r}_2)}P(\vec{r}_1,\vec{r}_2,
\vec{r}_3,\vec{r}_4)G(\vec{r}_3,\vec{r})G^*(\vec{r}\,^\prime,
\vec{r}_4)\nonumber\\
&&\left[ \delta  _{\hat{z}i}\delta  _{\hat{z}j}+
n_i n_j n_z^2-\delta _{\hat{z}i}n_jn_z-
\delta_{\hat{z}j}n_i n_z\right]
\end{eqnarray}
where $P$ is the diffusion propagator

\begin{equation}
\label{26}
P(\vec{r}_1,\vec{r}_2,\vec{r}_3,\vec{r}_4)=
\sum
\raisebox{-1.7cm}{
\setlength{\unitlength}{.3cm}
\begin{picture}(15,10)
\put(11,4){\vector(-1,0){10}}
\put(1,8){\vector(1,0){10}}
\multiput(1,4)(0,0.5){8}{\line(0,1){0.4}}
\multiput(3,4)(0,0.5){8}{\line(0,1){0.4}}
\multiput(5,4)(0,0.5){8}{\line(0,1){0.4}}
\multiput(7,4)(0,0.5){8}{\line(0,1){0.4}}
\multiput(9,4)(0,0.5){8}{\line(0,1){0.4}}
\multiput(11,4)(0,0.5){8}{\line(0,1){0.4}}
\put(1,2.5){\shortstack{$\vec{r}_2$}}
\put(1,8.5){\shortstack{$\vec{r}_1$}}
\put(11,2.5){\shortstack{$\vec{r}_4$}}
\put(11,8.5){\shortstack{$\vec{r}_3$}}
\end{picture}}
\end{equation}
Here the solid line denotes the averaged Green's function and the dotted one 
denotes the correlation function of random field. Respectively, the 
interference contribution has the form

\begin{eqnarray}
\label{27}
&& I^C_{ij}(\vec{R})=\frac{k ^2}{16\pi^2 R^2\varepsilon}\,
\int d\vec{r}d\vec{r}{\,}^\prime
B(r-r^\prime) A_0(\vec{r})A_0^*(\vec{r}{\,}^\prime)\,\int
d\vec{r}_1d\vec{r}_2 d\vec{r}_3 d\vec{r}_4\nonumber \\
&&e^{-ik\vec{n}(\vec{r}_1-\vec{r}_2)}P^{C}(\vec{r}_1,\vec{r}_2,
\vec{r}_3,\vec{r}_4)G(\vec{r}_3,\vec{r})G^*(\vec{r}{\,}^\prime,
\vec{r}_4)\nonumber\\
&&\left[ \delta  _{\hat{z}i}\delta  _{\hat{z}j}+
n_i n_j n_z^2-\delta _{\hat{z}i}n_jn_z-
\delta_{\hat{z}j}n_i n_z\right]
\end{eqnarray}
where

\begin{equation}
\begin{array}{c}
\label{28} P^C(\vec{r}_1,\vec{r}_2,\vec{r}_3,%
\vec{r}_4)= \sum%
\raisebox{-1.7cm}{
\setlength{\unitlength}{3mm}
\begin{picture}(15,10)
\put(11,4){\vector(-1,0){10}}
\put(1,8){\vector(1,0){10}}
\multiput(6,6)(-0.2,-0.4){6}{\circle*{.1}}
\multiput(6,6)(0.2,0.4){6}{\circle*{.1}}
\multiput(6,6)(0.2,-0.4){6}{\circle*{.1}}
\multiput(6,6)(-0.2,0.4){6}{\circle*{.1}}
\multiput(6,6)(.5,.2){10}{\circle*{.1}}
\multiput(6,6)(-.5,-.2){10}{\circle*{.1}}
\multiput(6,6)(-.5,.2){10}{\circle*{.1}}
\multiput(6,6)(.5,-.2){10}{\circle*{.1}}
\multiput(6,6)(0,1){8}{\line(-2,1){0.4}}
\multiput(6,6)(0,1){8}{\line(2,-1){0.4}}
\multiput(6,6)(0,1){8}{\line(-2,-1){0.4}}
\multiput(6,6)(0,1){8}{\line(-1,-2){0.4}}
\put(0,2.5){\shortstack{$\vec{r}_2$}}
\put(0,8.5){\shortstack{$\vec{r}_1$}}
\put(10,2.5){\shortstack{$\vec{r}_4$}}
\put(10,8.5){\shortstack{$\vec{r}_3$}}
\end{picture}}
\end{array}
\end{equation}
As follows from (\ref{28}), due to time-reversal invariance (see, for example 
[7]) $P^C$ is related to the diffusion propagator in the following manner

\begin{equation}
\label{29}
P^{C}(\vec{r_{1}},\vec{r_{2}},\vec{r_{3}},\vec{r_{4}})=P(\vec{r_{1}},
\vec{r_{4}},\vec{r_{3}},\vec{r_{2}})
\end{equation}
Thus the interference contribution as well as the diffusion one is determined 
by the diffusion propagator $P$. We now turn to finding it. Our consideration 
is similar to the three dimensional case [3]. It follows from (\ref{26}) that 
$P$ can be represented in the form

\begin{equation}
\label{30}
P(\vec{r}_1,\vec{r}_2,\vec{r}_3,\vec{r}_4)=B(\vec{r}_1-
\vec{r}_2)B(\vec{r}_3-\vec{r}_4)P(\vec{R}^\prime,\vec{r}_1
-\vec{r}_2,\vec{r}_3-\vec{r}_4)
\end{equation}
where $\vec{R}^\prime=\frac{1}{2}(\vec{r}_3+\vec{r}_4-\vec{r}_1-\vec{r}_2)$
and $P$ satisfies the equation

\begin{equation}
\label{31}
\int\frac{d\vec{p}}{(2\pi )^3}\left[ 1-\int\frac{d\vec{q}}
{(2\pi )^3}f(\vec{q},\vec{K})B(\vec{p}-\vec{q})\right]
P(\vec{K},\vec{p},\vec{q}{\,}^\prime)=f(\vec{q}{\,}^\prime,\vec{K})
\end{equation}
where

\begin{equation}
\label{32}
f(\vec{q},\vec{K})=G(\vec{q}+\vec{K}/2)G^*(\vec{q}-\vec{K}/2)
\end{equation}
As usual, one needs to know $P$ at $K\rightarrow 0$. In this limit the 
diffusion propagator has the form [3]

\begin{equation}
\label{33}
P(\vec{K}\rightarrow 0,\vec{p},\vec{q})=\frac{{\rm Im}G(\vec{p})
{\rm Im}G(\vec{q})}{{\rm Im}\Sigma(\vec{q})}A(\vec{K})
\end{equation}
where

\begin{equation}
\label{34}
A(\vec{K})=\left[ 3\int\frac{(\vec{q}\vec{K})^2 {\rm Im}G(\vec{q})}
{{\rm Im}^2\Sigma(\vec{q})}\,\frac{d\vec{q}}{(2\pi )^3}\right]^{-1}
\end{equation}
Substituting (\ref{13}) and (\ref{14}) into $(\ref{34})$ and calculating 
the integral in the limit $ka\gg 1$, we obtain

\begin{equation}
\label{35}
A(\vec{K})=\frac{20\pi}{k}\,\frac{1}{3K_{z}^2 l^2+K_{\rho}^2 l^2}
\end{equation}
In our previous paper [2] we investigated the special case $K_{\rho}=0$, which 
was sufficient for studying the diffusion contribution. Note that although we 
considered the case $ka\gg 1$ all results are qualitatively correct also in 
the in the general case. 
When we know the form of diffusion propagator we 
can investigate the
investigate the diffusion and interference contributions. First we consider 
the diffusion contribution. Substituting (\ref{30}) into (\ref{25}) and 
using (\ref{33}), (\ref{23}) and (\ref{14}), we obtain [2]

\begin{eqnarray}
\label{36}
&&I^D(\vec{n})=\frac{e ^2}{2c \varepsilon}(1-n_z^{2})
A(\vec{K})k^2{\rm Im} \Sigma(k\vec{n})L_z\times\nonumber \\
&\times&\int\frac{d\vec{q}_\rho }{(2\pi )^2}
\frac{1}{(q^2_{\rho}+k_0^2-k^2)^2}
\frac{\left[ B\left( |k_0+\sqrt{k^2-q_\rho ^2}|\right)+
B\left( |k_0-\sqrt{k^2-q_\rho ^2}|\right)\right]}
{B(0)+B(2\sqrt{k^2-q_{\rho}^2)}}
\end{eqnarray}
  As seen from (\ref{36}), the main contribution to the integral over 
$q_{\rho}$ for relativistic particles $\gamma\gg 1$, $k_0\rightarrow k$ give
the values $q_{\rho}\approx 0$. Accounting for this circumstance and using
(\ref{35}) we have from (\ref{36})

\begin{equation}
\label{37}
I^D(\omega, \theta)=\frac{5}{6}\,\frac{e^2\gamma
^2}{\varepsilon c}\,\left( \frac{L_z}{l(\omega)}\right)^3\,\frac{\sin^2\theta}{|\cos\theta|}
\end{equation}
  When obtaining  (\ref{37}) we substitute as usually $1/K^2$ at $K\rightarrow
0$ by $L_{z}^2$ (there with we assume that $L_{z}<L_{x},L_{y}$).
% Note the  main peculiarities of the diffusion contribution.
A more precise approach would be to solve the appropriate
Schwarzschild-Milne equation for the present problem. This could bring
overall prefactors of order unity [8,4]. For our present purpose we
shall not be interested in these effects.
   It follows from  (\ref{37}) and  (\ref{24}), that $I^D/I^0\sim (L_{z}/l)^2
\gg 1$. Thus when $k|\cos{\theta}|l\gg 1$ and $l\ll L_z$ the diffusion 
contribution is superior to the single scattering one. As one could expect, 
the forward and backward intensities in the diffusion contribution are equal 
to each other.
   Finally, we note the strong dependence of the
diffusion contribution on the particle  energy.
 
\section{Interference Contribution}

   Using  (\ref{29}) and (\ref{30}) and changing the variables of integration 
by formulae

\begin{equation}
\label{38}
\vec{x_1}=\vec{r_1}-\vec{r_4}, \vec{x_2}=\vec{r_3}-\vec{r_2},
\vec{R{\,}^{\prime}}=\frac{1}{2}(\vec{r_3}+\vec{r_2}-\vec{r_1}-\vec{r_4}),
\vec{r_4}\equiv\vec{r_4}
\end{equation}
we find from (\ref{27})

\begin{eqnarray}
\label{39}
&& I^C(\vec{n})=\frac{(1-n_{z}^{2})ck ^2}{32\pi^2 \varepsilon}\,
\int d\vec{r}d\vec{r}{\,}^\prime
B(r-r^\prime) A_0(\vec{r})A_0^*(\vec{r}{\,}^\prime)\,\int
d\vec{x}_1d\vec{x}_2 d\vec{R}{\,}^{\prime} d\vec{r}_4\nonumber \\
&&e^{ik\vec{n}(\vec{R}-\frac{\vec{x_1}+\vec{x}_2}{2})}
P(\vec{R}{\,}^{\prime},\vec{x}_1,
\vec{x}_2)B(\vec{x}_1)B(\vec{x}_2)G(\vec{R}{\,}^{\prime}+\frac{\vec{x_1}+\vec{x}_2}{2}+\vec{r_4}-\vec{r})G^*(\vec{r}{\,}^{\prime}-\vec{r}_4)
\end{eqnarray}
   In the Fourier representation one finds from (\ref{39})

\begin{eqnarray}
\label{40}
I^C(\vec{n})=\frac{(1-n_{z}^{2})ck ^2}{32\pi^2 \varepsilon}\,
\int \frac{d\vec{q}d\vec{q_1}d\vec{q_2}d\vec{K}}{(2\pi)^{12}}
B(\vec{q})|A_0(-k\vec{n}-\vec{K}-\vec{q})|^2\nonumber \\
B(\vec{q}_1)B(\vec{q}_2)P(\vec{K},\vec{k},
\vec{n}+\frac{\vec{K}}{2}-\vec{q_1},k\vec{n}+\frac{\vec{K}}{2}-\vec{q_2})
|G(k\vec{n}+\vec{K})|^2
\end{eqnarray}
   Substituting (\ref{13}),     (\ref{23}),  (\ref{33})   and  (\ref{34})   
into (\ref{40}), taking into account that the main contribution in the 
integral over $\vec{K}$ give the values $K\rightarrow 0$ and
   sequentially integrating (\ref{40}) using the Ward identity, we find from
(\ref{40})

\begin{eqnarray}
\label{41}
I^C(\vec{n})=\frac{10\pi e^2 L_z |n_z|(1-n_z)^2B(|k\cos{\theta}+k_0|)}
{\varepsilon c l}\,
\int \frac{d\vec{K}}{(2\pi)^{3}}\frac{1}{(3K_z^2+K_{\rho}^2)\left[(\vec{K_\rho}
+k\vec{n_\rho})^2+k_0^2\gamma^{-2}\right]^2}
\end{eqnarray}
   Note that at $ak\gg 1$, $B(2k)/B(0)\sim 1/k^2a^2\ll 1$. Therefore, as 
follows from (\ref{41}) in the Cherenkov limit the backward intensity 
($\theta\approx\pi$) is significantly larger than 
   the forward intensity ($\theta\approx 0$). This is the main characteristic 
feature of the interference contribution. It is analogous to the light back-
   scattering peak which occurs in propagation of light in the randomly
   inhomogeneous media [6].
    Considering angles  $\sin{\theta}\gg\lambda/2\pi L_{\rho}$ and calculating 
the integral over $K$ in (\ref{41}) we have

\begin{equation}
\label{42}
I^C(\theta)=\frac{5 \sqrt{2}\arctan{\sqrt{2}}}{2\pi\varepsilon}\frac{e^2}{c}
\frac{L_z}{l^2}\frac{B(|k\cos{\theta}+k_0|)\sin^2{\theta}|\cos{\theta}|}
{\left[k^2\sin^2{\theta}+k_0^2\gamma^{-2}\right]^2}
\end{equation}
    When obtaining (\ref{42}) we cut the integral over $K$ on the upper limit 
at 1/l.
    Note that the ratio $\lambda/2\pi L_{\rho}$, in the optical region, is of 
order $\sim 10^{-4}-10^{-5}$. Therefore one can believe that the condition 
$\sin{\theta}\gg\lambda/2\pi L_{\rho}$ is always satisfied in the optical 
region. Comparing (\ref{42}) with the single scattering contribution
    (\ref{24}) we see that $I^{C}/I^{0}\sim 1/(kl)^{2}\ll 1$. Although the 
interference contribution is small however it has quite different angular 
dependence. 
  Accounting for the form of correlation function $B$ (\ref{16}), it follows 
from (\ref{42}) that the maximum of the radiation intensity for relativistic 
particles $\gamma\gg 1$, $k_0\rightarrow k$, lies in the region of angles
$\theta\sim\pi-\gamma^{-1}$. These angles
are very close to the backward direction. On the other hand, the
  maximum of the single scattering contribution lies in the 
strongly forward range of angles
$\theta\sim\gamma^{-1}$.

\section{Summary}

   We have considered the influence of interference effects on the radiation
   of a charged particle passing through a stack of randomly spaced plates.
   It appears that interference contribution to the radiation intensity has
   a peak in the backward to particle motion direction. Though its value is
   small compared to the single scattering and diffusion
contributions, it can be investigated experimentally. This is possible
 due its specific angular dependence.

\end{document}